\documentclass[a4paper,aps,twocolumn,amsmath,amssymb,showpacs]{revtex4}
\usepackage{graphicx}
\usepackage{graphics}

\newcommand{\bsigma}{\mbox{\boldmath ${\sigma}$}}

\newcommand{\odis}{\big<\hspace{-1mm}\big<}
\newcommand{\cdis}{\big>\hspace{-1mm}\big>}
\newcommand{\Odis}{\Big<\hspace{-1.5mm}\Big<}
\newcommand{\Cdis}{\Big>\hspace{-1.5mm}\Big>}
\newcommand{\Odiss}{\Bigg<\hspace{-2mm}\Bigg<}
\newcommand{\Cdiss}{\Bigg>\hspace{-2mm}\Bigg>}

\begin{document}
\title{Gardner optimal capacity of the diluted Blume-Emery-Griffiths 
neural network}
\author{D. Boll\'e}
\email{Desire.Bolle@fys.kuleuven.ac.be}
\author{I. P\'erez Castillo}
\email{Isaac.Perez@fys.kuleuven.ac.be}
\affiliation{Instituut voor Theoretische Fysica, 
Katholieke Universiteit Leuven,  B-3001 Leuven, Belgium}
\begin{abstract}
The optimal capacity of a diluted Blume-Emery-Griffiths neural network 
is studied as a function of the pattern activity and the embedding
stability using the Gardner entropy approach. 
Annealed dilution is considered, cutting some of the couplings referring 
to the ternary patterns themselves and some of the couplings 
related to the active patterns, both simultaneously (synchronous dilution)
or independently (asynchronous dilution). 
Through the de Almeida-Thouless criterion it is found that the 
replica-symmetric solution is locally unstable as soon as there is dilution.
 The distribution of the couplings shows the typical 
gap with a  width depending on the amount of dilution, but this gap 
persists even in cases where a particular type of coupling plays no 
role in the learning process. 
\end{abstract}
\maketitle
\section{Introduction}
\label{sec:Introduction}
During recent years the Blume-Emery-Griffiths (BEG) model \cite{BEG} has 
been studied quite intensively in the context of neural networks, one of
the reasons being that it was argued in \cite{Do00} that this model
maximizes the mutual information content of three-state networks with
Hebbian-type learning rules. 
To know in more detail how the retrieval quality of the BEG network
compares with other three-state neuron models, the thermodynamics of
this model was studied and temperature-capacity phase diagrams were
obtained \cite{Ve03}. It was shown that the retrieval phase is 
systematically larger than that of other three-state models and that 
the critical capacity is about twice as large as that of the 
three-state neuron Ising model \cite{BRS94}. Also the region of
thermodynamic stability is much larger and, furthermore, the phase 
diagram itself is much richer with the presence of a stable quadrupolar 
state, carrying also retrieval information, at high temperatures.  

It was also shown that this enhancement of the retrieval properties is
not restricted to the use of the Hebbian learning rule but that it is
inherent to the model. Indeed, by studying the
Gardner optimal capacity \cite{Ga87} in replica symmetric (RS) mean-field 
theory it was found recently \cite{Pe03} that for the corresponding BEG 
perceptron with, e.g.,  zero embedding stability parameter   
and uniform patterns this capacity is $2.24$.
Comparing with other three-state neuron perceptron models, we recall
that for the $Q=3$ Ising perceptron the Gardner optimal capacity can 
maximally reach $1.5$ \cite{MKB91, BDM91},
 whereas for the $Q=3$ clock and Potts 
model both reach an optimal capacity of $2.40$ \cite{GK94, GBK}. 
At this point we have to remark that the $Q=3$ Ising perceptron and the 
BEG perceptron have the same topology structure in the neurons, whereas
the $Q=3$ clock and Potts models have different topologies.
For the Ising topology structure the BEG-perceptron has the best 
performance.

The interesting question remains whether and in how far these enhanced 
retrieval properties are robust against dilution. Studying this question is the
aim of the present work. Besides the fact that the
connectivity of biological networks is far from complete, diluted networks 
offer the possibility to
study the robustness against malfunctioning of some of the connections.
Furthermore, in asymmetric architectures they reduce the internal 
feedback correlations of fully
connected networks making a complete analytic description of the dynamics
much easier \cite{Do00}, \cite{De87, Ga89}. Finally, in the BEG
perceptron there are two sets of couplings, those referring to
the three-state patterns themselves and those related to the
active, i.e., the non-zero patterns. By diluting both types of couplings 
simultaneously or  diluting these couplings independently, we can study, in
particular,  the influence of 
the active patterns on the Gardner optimal capacity of the BEG
perceptron. These results can be obtained in closed analytic form. 

We remark that the type of dilution  we study in this paper is such
that the number of connections to a given site still increases with the 
size of the system. In the replica approach to capacity problems for
these systems, only order parameters with two replica indices appear.
Recently, the study of neural networks with finite connectivity, i.e.,
where the number of connections to a given site remains finite in the
thermodynamic limit has been started \cite{Co03, Pe03b}. There,
functional order parameters have to be introduced.

The paper is organized as follows. In  Sect.~\ref{sec:definition} we 
recall the BEG model and briefly discuss  some of its properties. In 
Sect.~\ref{sec:Gardner} we introduce the different kinds of dilution
that one may study and report on the application of the Gardner 
approach to these cases. We present the results for the optimal 
capacity in the RS approximation as a function of the pattern 
activity, the stability parameter and the degree of dilution. In 
Sect.~\ref{sec:distribution} we discuss the results  for the distribution
of the couplings and in Sect.~\ref{sec:stability} we study the validity
of the local stability criterion for the RS solution. The last 
section contains the conclusions.

\section{The BEG neural network}
\label{sec:definition}
Let us consider a neural network consisting of $N$ neurons which can 
take values $\sigma_i, i=1,\ldots, N$ from the discrete set 
${\mathcal S}\equiv\lbrace -1,0,+1 \rbrace $. The $p$ patterns 
to be stored in this network are supposed to be a collection 
of independent and identically distributed random 
variables (i.i.d.r.v.), ${\xi}_i^\mu $, $\mu=1,\ldots,p$, taken from the set 
${\mathcal S}$ with a probability distribution
\begin{equation}
p(\xi_{i}^{\mu})=\frac{a}{2}\delta(\xi_{i}^{\mu}-1)
+\frac{a}{2}\delta(\xi_{i}^{\mu}+1)+(1-a)\delta(\xi_{i}^{\mu})
\label{distribution}
\end{equation}
with $a$ the activity of the patterns so that
\begin{equation}
 \lim_{N\rightarrow\infty}\frac{1}{N}\sum_{i}(\xi_{i}^{\mu})^2 = a.
\end{equation}
Given the network configuration at time 
$t$, ${\bsigma}_N\equiv\{\sigma_j(t)\}, j=1,\ldots,N$,
 the following dynamics is considered. The configuration 
$\bsigma_N(0)$ is chosen as input. The neurons are updated 
according to the stochastic parallel spin-flip dynamics 
defined by the transition probabilities
\begin{eqnarray}
\Pr \left(\sigma_i(t+1) = s' \in {\mathcal S}| \bsigma_N(t) \right) 
       \nonumber   \\  && \hspace*{-1.9cm}
       = \frac
        {\exp [- \beta \epsilon_i(s'|\bsigma_N(t))]}
        {\sum_{s \in \mathcal{S}} \exp [- \beta \epsilon_i
                                   (s|\bsigma_N(t))]}\,.
     \label{eq:trans}
\end{eqnarray}
Here the energy potential $\epsilon_i[s|{\bsigma}_N(t)]$ is
 defined by
\begin{equation}
          \epsilon_i[s|{\bsigma}_N(t)] =
          -sh_i({\bsigma}_N(t))-s^2\theta_i({\bsigma}_N(t))
              \,,
\label{eq:energy}
\end{equation}
where the local fields in neuron $i$, 
$h_{N,i}(t)\equiv h_i({\bsigma}_N(t))$ carry all the information
\begin{equation}
        \label{eq:h}
      h_{N,i}(t)=\sum_{j \neq i} J_{ij}\sigma_j(t), \quad
      \theta_{N,i}(t)=\sum_{j\neq i}K_{ij}\sigma_{j}^{2}(t)\, .
\end{equation}
At zero temperature the updating rule  of this dynamics  
(\ref{eq:trans})-(\ref{eq:energy}) is equivalent to the gain
function formulation
\begin{eqnarray}
       \sigma_i(t+1)&=&\mbox{sign}(h_{N,i}(t)) \Theta(|h_{N,i}(t)| 
       + \theta_{N,i}(t))
       \nonumber \\
       &\equiv& \mbox{g}(h_{N,i}(t), \theta_{N,i}(t))
\label{updating_rule}
\end{eqnarray}
with $\Theta(x)$ and  $\mbox{sign}(x)$ the Heaviside and the 
sign function, respectively.  

Concerning the loading capacity of this model, the following results
have appeared in the literature. For Hebbian-type synaptic couplings $J_{ij}$ 
and $K_{ij}$ 
\begin{eqnarray}
&&J_{ij}=\frac{1}{a^{2}N}\sum_{\mu=1}^{p}\xi_{i}^{\mu}\xi_{j}^{\mu}
      \label{hamcoupj}     \\
&&K_{ij} = \frac{1}{a^2(1-a)^2N} \sum_{\mu=1}^p
         [(\xi^\mu_i)^2 - a] [(\xi^\mu_j)^2 -a]
     \label{hamcoupk}
\end{eqnarray}
the long-time behavior is governed by the Hamiltonian
\begin{equation}
H = - \frac{1}{2} \sum_{i \neq j} J_{ij} \sigma_i \sigma_j
    - \frac{1}{2} \sum_{i \neq j} K_{ij} \sigma_i^2 \sigma_j^2 \ ,
 \label{hamil}
\end{equation}
and the retrieval properties are enhanced \cite{Ve03} in comparison to
other three-state neuron models. In particular, the retrieval phase is 
systematically larger than that of other three-state models and  
the critical capacity is about twice as large as that of the 
three-state neuron Ising model. Moreover, depending on the value of 
the pattern activity a stable quadrupolar  state carrying also non-zero
retrieval information arises at high temperatures. 
However, an underlying reason why there is such an enlargement 
of the basin of attraction and hence of the retrieval 
properties of the network seems still to be absent.

This enhancement of retrieval has also been found \cite{Pe03} for the 
BEG-perceptron
 \begin{equation}
\xi^\mu_0=\text{ sgn}(h^\mu)\Theta(|(h^\mu|+\theta^\mu),
\quad \forall \mu=1,\ldots,p
\label{restriction}
\end{equation}
with $\xi^\mu_0$ denoting the output, and where $h^\mu$ and $\theta^\mu$
are the local fields at the output created by the pattern $\mu$
\begin{equation}
h^\mu=\frac{1}{\sqrt{ N}}\sum_{i=1}^NJ_{i}\xi^\mu_i \, ,\quad 
\theta^\mu=\frac{1}{\sqrt{ N}}\sum_{i=1}^N K_{i}(\xi^\mu_i)^2
\label{local_fields}
\end{equation}
with $ J_i, K_i$ a set of couplings connecting the input 
with the output. In a RS analysis the Gardner optimal 
capacity for this perceptron is calculated analytically and seen to be
bigger than that of the $Q=3$-Ising perceptron~\cite{MKB91, BDM91}.

\section{The diluted BEG perceptron}
\label{sec:Gardner}
We want to find out in how far these enhanced retrieval properties are
robust against dilution. One of the questions we want to answer then is 
the following. Let 
$\xi^\mu_i, {\mu=1,\ldots,p},{i=0,\ldots,N}$ be an 
extensive set of $p=\alpha N$ patterns supposed
to be fixed points of the dynamical rule \eqref{restriction}
where the local fields $h^\mu$ and $\theta^\mu$ are now given by 
\begin{equation}
h^\mu=\frac{1}{\sqrt{c_J N}}\sum_{i=1}^Nc^J_{i}J_{i}\xi^\mu_i \, ,\quad 
\theta^\mu=\frac{1}{\sqrt{c_K N}}\sum_{i=1}^Nc^K_iK_{i}(\xi^\mu_i)^2\, .
\label{local_fields2}
\end{equation}
The parameters $c^J_{i}\in\{0,1\}$ and 
$c^K_i\in\{0,1\}$ control the presence of the connections $J_i$ and 
$K_i$. We want to find a  set of couplings, $ J_i^\star, K_i^\star$,
or equivalently, a BEG-perceptron with an average dilution 
$c_J$ and $c_K$ 
\begin{equation}
  c_J=\frac{1}{N}\sum_{i=1}^N c^J_{i},
           \quad c_K=\frac{1}{N}\sum_{i=1}^N c^K_{i}
\label{constraint1}
\end{equation}
that still fulfil the conditions \eqref{restriction}. It is clear that
for small values of the capacity $\alpha$  more than one BEG-perceptron
storing these  patterns can be found. The bigger the value of $\alpha$ 
the more difficult this task becomes and a saturation limit, 
called Gardner optimal capacity, is reached.

In the following we study dilution during learning, i.e., annealed dilution, 
which can be realized in two different ways. The first one, called synchronous 
dilution, assumes that $c^J_i=c^K_i\equiv c_i$ and, hence $c_J=c_K$; the
second one, named asynchronous dilution, allows the $c_i$'s to be
different.  Looking back at \eqref{local_fields2}  we see that the $c^J_i$
control the connections of the three-state patterns, while the $c^K_i$
control the connections related to the active, i.e., the non-zero
patterns. In fact, for Hebbian learning  in 
\eqref{hamcoupj}-\eqref{hamil}  the $K$-couplings 
control the fluctuations around these active patterns. Therefore, by allowing
synchronous or asynchronous dilution we can study the influence of 
the active patterns on the  optimal capacity of the BEG
perceptron.

\subsection{Synchronous dilution}
To study the optimal capacity, we follow the entropy approach introduced 
by Gardner \cite{Ga87}. Since the dynamical variables are continuous, 
entropy has only meaning relatively and we write the volume $V$ of all
possible BEG-perceptrons satisfying \eqref{restriction}, without 
normalizing, as
\begin{equation}
V=\prod_{i=1}^N\underset{c_i, J_i,K_i}{\text{tr}}
\prod_{\mu=1}^p\chi_{\xi^\mu_0}(h^\mu,\theta^\mu;\kappa)
\end{equation}
with $\chi_{\xi^\mu_0}(h^\mu,\theta^\mu;\kappa)$ the characteristic 
function given by 
\begin{eqnarray}
\chi_{\xi^\mu_0}(h^\mu,\theta^\mu;\kappa)
&=&(\xi^\mu_0)^2\Theta(\xi^\mu_0 h^\mu-\kappa)
\Theta(|h^\mu|+\theta^\mu-\kappa)\nonumber\\
&&+[1-(\xi^\mu_0)^2]\Theta(-|h^\mu|-\theta^\mu-\kappa)
\label{charateristic_function}
\end{eqnarray}
where  $\kappa$ is the embedding stability parameter.  Since we consider
continuous couplings we need to introduce a modified spherical constraint 
\begin{equation}
\sum_{i=1}^Nc_iJ^2_{i}=cN,\quad \sum_{i=1}^Nc_iK^2_{i}=cN\,.
\label{constraint23}
\end{equation}
From this spherical constraint we see that the couplings are not well 
normalized at those sites where $c_i$ is zero. One can solve this
difficulty either by introducing an extra spherical constraint for 
the remaining couplings \cite{Bo90}, either by restricting the trace 
over the couplings  \cite{Mo97}. We take the  second solution and define 
the restricted trace as
\begin{equation}
\underset{ c_i, J_i, K_i}{\text{tr}}(\cdots)
   \equiv \sum_{c_i=0,1}\delta_{c_i,0}(\cdots)
      +\sum_{c_i=0,1}\delta_{c_i,1}\int dJ_idK_i(\cdots)\, .
\end{equation}
Since we want to study typical features of the system the important quantity 
to average over is the entropy.  Employing replica techniques \cite{MPV} we 
express the entropy per neuron as
\begin{equation}
  v=\lim _{N \to \infty}\lim_{n\to 0}\frac{1}{n N}\ln\odis V^n\cdis
\label{entropybis}
\end{equation}
where $\odis \cdot \cdis$ denotes the average over the pattern distribution
\eqref{distribution} and where  $V^n$ is the $n$-th times replicated
volume of solutions
\begin{widetext}
\begin{equation}
 V^n=\Big[\prod_{j=1}^N\prod_{\alpha=1}^n
     \underset{ c_i^\alpha,J_i^\alpha,K_i^\alpha}{\text{tr}}
     \Big]   \Big[\prod_{\alpha=1}^n\delta\Big(\sum_{i=1}^N 
          c^\alpha_{i}(J^\alpha_i)^2-cN\Big)
          \delta\Big(\sum_{i=1}^N c^\alpha_{i}(K^\alpha_i)^2-cN\Big)
         \delta\Big(\sum_{i=1}^N c^\alpha_{i}-cN\Big)\Big]
       \prod_{\mu=1}^p\prod_{\alpha=1}^n
             \chi_{\xi^\mu_0}(h^\mu_\alpha,\theta^\mu_\alpha;\kappa).
   \label{power_volume}
\end{equation}
\end{widetext}
The further analysis then proceeds in a standard way although the
technical details are much more involved. A short account 
 is given in Appendix~A. 

The results are described essentially in terms of three order
parameters, the first one, $q_{\alpha \beta}$, defined as the overlaps 
between two distinct replicas for the couplings $J_i$, the second one, 
$r_{\alpha \beta}$, a similar quantity for the couplings $K_i$ and the third
 one, $L^\alpha$, arising from the fact that the
dynamics and, hence, also the characteristic function contains a second
field $\theta$, quadratic in the patterns (see \eqref{order_parameters}).
In the RS approximation we are discussing here they are 
given by    
$q_{\alpha\beta}=q, r_{\alpha\beta}=~r,L^\alpha=L$.

The RS  Gardner optimal capacity is obtained when the overlap order
parameters $q$ and $r$ go to $1$. It is clear that these limits have to
be taken simultaneously but, in general, their rate of convergence could
be different. Therefore, we introduce $(1-r)=\gamma(1-q)$ where $\gamma$
is a new parameter which one also needs to extremize. We expect this
parameter $\gamma$ to depend on the pattern distribution through the
activity $a$. The result for the replica symmetric Gardner optimal capacity 
$\alpha^{RS}_{\text{syn}}$ then reads  
\begin{equation}
\alpha^{{RS}}_{\text{syn}}(a,\kappa,c)=
   \underset{u,L,\gamma}{\text{extr}}\,\,\,
        \frac{A_{\text{syn}}(u,\gamma;c)}{g(\gamma,L;a,\kappa)}
   \label{capacity_RS_synchronous}
\end{equation}
where the function $ A_{\text{syn}}(u,c;\gamma)$ is defined by
\begin{equation}
   A_{\text{syn}}(u,\gamma;c)=
     u^2 c +  A^{(2)}_{\text{syn}}(u,\gamma)\,.
\end{equation}
Stationarity with respect to $u$  then leads to
$c=A^{(1)}_{\text{syn}}(u,\gamma)$. Here the functions $A^{(m)}_{\text{syn}}$,
 $m=1,2$ are given by 
\begin{equation}
A^{(m)}_{\text{syn}}(u,\gamma)=\frac{m\sqrt{\gamma}}{2\pi}\int_{0}^{2\pi}
d\varphi\frac
  {\exp[{-\frac{u^2}{2}\big(\cos^2\varphi+\gamma\sin^2\varphi\big)}]}
   {\big(\cos^2\varphi+\gamma\sin^2\varphi\big)^m}.
\end{equation}

The function $g(\gamma,L;a,\kappa)$ in \eqref{capacity_RS_synchronous}
can be expressed as 
\begin{widetext}
\begin{eqnarray}
g(\gamma,L;a,\kappa)&=& 
                a\sum_{i=1}^3\int_{{\cal R}_{i}}
   {\cal D}(h_0+\kappa/\sqrt{a}){\cal D}(\sqrt{\gamma}\theta_0-l_0)
               d_{min}^{{\cal R}_{i}}(h_0,\theta_0)
	         \nonumber\\
&&+(1-a)\sum_{i=1}^3\int_{{\cal R}'_{i}}{\cal D}(h_0) 
    {\cal D}(\sqrt{\gamma}\theta_0-l_{\kappa})
                       d_{min}^{{\cal R}'_{i}}(h_0,\theta_0)
                     \nonumber\\
    \label{grs}
\end{eqnarray}
with  $l_{\kappa}\equiv (aL+\kappa)/\sqrt{a(1-a)}$
and ${\cal D}(ax +b) = (2\pi)^{-1/2}a \exp[(-1/2)(ax+b)^{2}]\, dx$.

The integration regions are the following ones
\begin{eqnarray}
{\cal R}_1=\left\{\begin{array}{l}
h_0<0\\
\theta_0>0
\end{array}\right.\label{Region1}&
{\cal R}_2=\left\{\begin{array}{l}
h_{0}\gamma'<\theta_{0}<0\\
h_{0}<0
\end{array}\right.\label{Region2}&
{\cal R}_3=\left\{\begin{array}{l}
\theta_{0}<0\\
\theta_{0}/\gamma'<h_{0}<-\theta_{0}\gamma'\\
\end{array}\right.\label{Region3}\\
{\cal R}'_1=\left\{\begin{array}{l}
h_{0}>0\\
-h_0/\gamma'<\theta_{0}<\gamma' h_{0}
\end{array}\right.\label{Region1prim}&
{\cal R}'_2=\left\{\begin{array}{l}
-\theta_0/\gamma'<h_{0}<\theta_0/\gamma'\\
\theta_{0}>0
\end{array}\right.\label{Region2prim}&
{\cal R}'_3=\left\{\begin{array}{l}
h_{0}<0\\
h_0/\gamma'<\theta_{0}<-\gamma' h_{0}
\end{array}\right.\label{Region3prim}
\end{eqnarray}
and the corresponding integrands are given by
\begin{eqnarray}
d^{{\cal R}_1}_{min}=h_0^2\label{Mindis1}\quad\quad&
d^{{\cal R}_2}_{min}=h_0^2 +\theta_0^2\label{Mindis2}\quad&
d^{{\cal R}_3}_{min}=\frac{1}{1+(\gamma')^2}\big(h_0+ 
        \gamma'\theta_0\big)^2\label{Mindis3}\\
d^{{\cal R}'_1}_{min}=\frac{1}{1+(\gamma')^2}\big(h_0+
         \gamma'\theta_{0}\big)^2\label{Mindis1prim}\quad&
d^{{\cal R}'_2}_{min}=h^2_{0}+\theta_{0}^2\label{Mindis2prim}&
d^{{\cal R}'_3}_{min}=\frac{1}{1+(\gamma')^2}\big(h_0-
        \gamma'\theta_{0}\big)^2\label{Mindis3prim}
\end{eqnarray}
\end{widetext}
with $\gamma'\equiv \sqrt{\gamma(1-a)}$. 

After inserting \eqref{grs}-\eqref{Mindis3prim} in 
\eqref{capacity_RS_synchronous} and extremizing numerically 
we find the results presented in Fig.~1. We plot the  optimal capacity 
$\alpha(a,\kappa,c)$ itself (insets) and its values normalized by the 
optimal capacity for no dilution, $\alpha(a,\kappa,c)/\alpha(a,\kappa,1)$, 
as a function of  the dilution $c$ for $\kappa=0$ and several values of 
the activity $a$.
\begin{figure}[ht]
\begin{center}
\includegraphics[width=0.30\textwidth,height=0.25\textheight]
{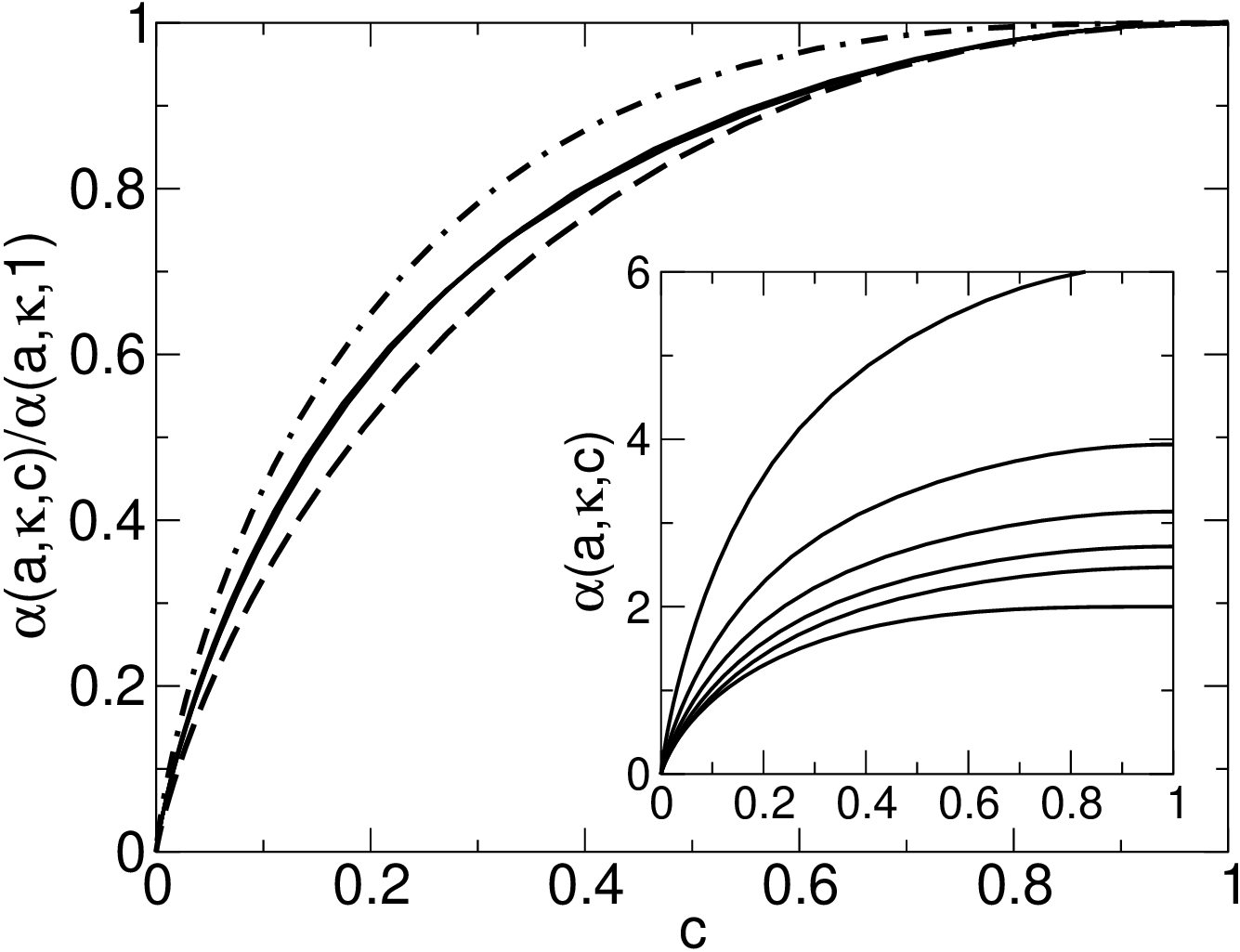}
\\
\vspace{1cm}
\includegraphics[width=0.30\textwidth,height=0.25\textheight]
{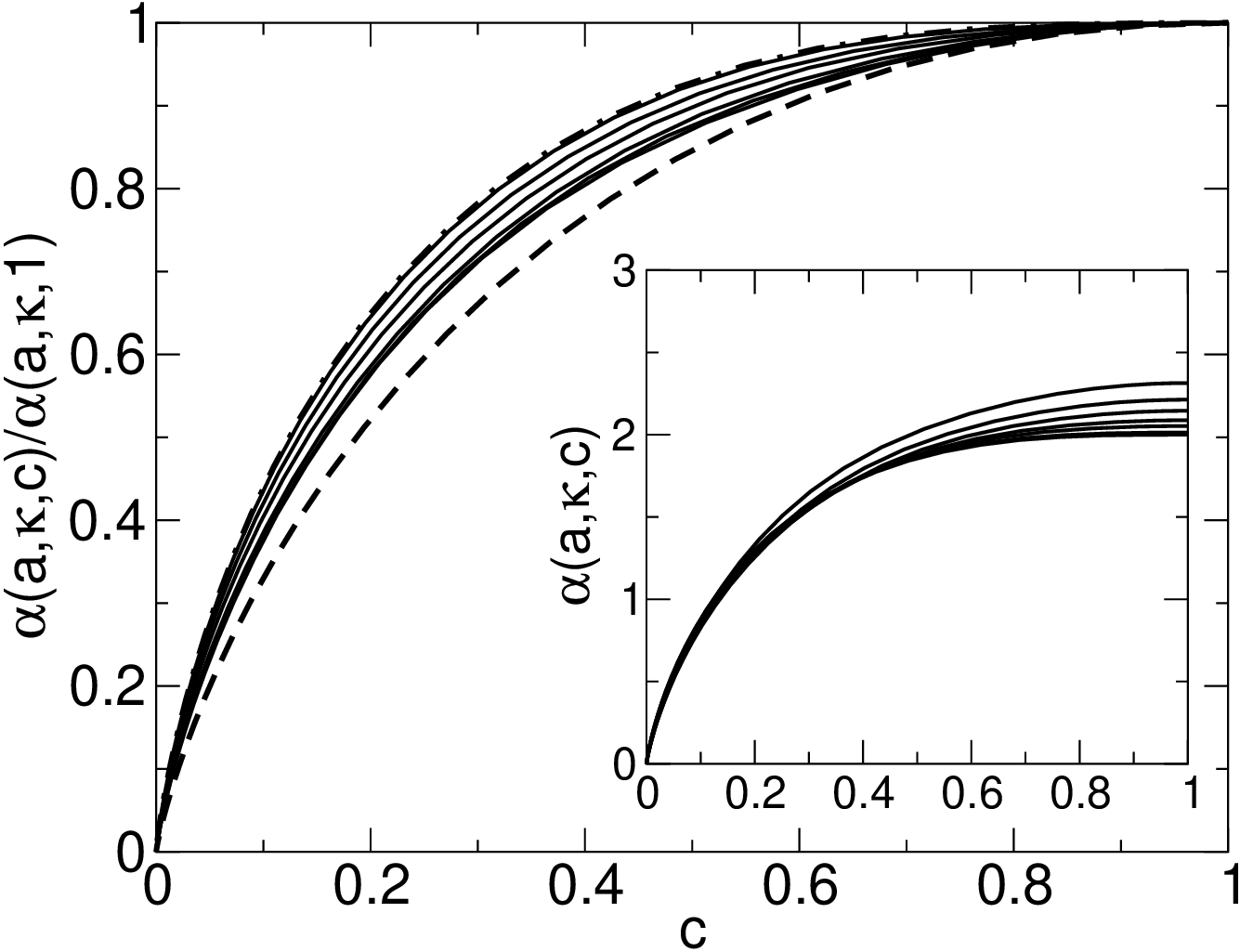}
\caption{Optimal capacity for synchronous dilution in the 
BEG-perceptron as a function of $c$ for $\kappa=0$ and several values of
$a$. 
Top figure: the normalized optimal capacity (solid line), its upper
bound (dashed-dotted line) and its lower bound (broken line); 
the inset displays $\alpha(a,\kappa,c)$ for $a=0.1,0.2,0.3,0.4,0.5$ from  
top to bottom.  
Bottom figure : similar to the top figure for 
 $a=0.6,0.7,0.8,0.9,0.95,0.99$ .}
\end{center}
\end{figure}

We see that different regions of activities lead to different results.
For small activities $a \leq 0.5$ and hence, many inactive neurons, the 
optimal 
capacity strongly increases for decreasing dilution. This seems to be
in agreement with what is known in the literature for very small
activities or so-called sparse coding (see, e.g., \cite{T88},
\cite{PV89}, \cite{H89}). When normalizing these results by
$\alpha(a,\kappa,1)$ we find that all the lines collapse into the
full line.
For large  activities $a \geq 0.6$ and, hence, many active
states $\pm 1$, the results for $\alpha(a,\kappa,c)$  are only weakly
dependent on the activity (see inset) but the results for the normalized
optimal capacity do not  collapse. Furthermore, we see that the network is
more robust against synchronous dilution for non-sparse coding, i.e.,
for activities ranging in the interval $[\,0.2, 1.0\,]$: the dilution $c$
can decrease from $1$ to about $0.4$ before one sees a substantial
decrease in the optimal capacity. Comparing to the $Q=3$ Ising
perceptron \cite{Mo97}, the effect of dilution, especially for larger
activities is about the same.

When $c=1$ ($u=0$), the functions $A^{(m)}_{\text{syn}}(u,\gamma)$
 can be explicitly integrated leading to
\begin{equation}
A^{(1)}_{\text{syn}}(0,\gamma)=1,\qquad A^{(2)}_{\text{syn}}(0,\gamma)
=1+\frac{1}{\gamma}
\end{equation}
and one  recovers the optimal capacity found in the fully connected 
case \cite{Pe03}. When the pattern activity $a$ goes to $1$ 
the system is forced into two possible states, as in the Gardner
model with dilution \cite{Bo90}. Since the overlap parameter $r$ becomes
irrelevant in such a limit $\gamma$  must go to infinity. The numerical
solution does confirm this. Furthermore, in this limit the functions 
$A^{(m)}_{\text{syn}}(u,\gamma)$   become
\begin{eqnarray}
&&A^{(1)}_{\text{syn}}(u,\infty)=\text{erfc}\Big(\frac{u}{\sqrt{2}}\Big)
 \nonumber \\
&&A^{(2)}_{\text{syn}}(u,\infty)=\frac{2u}{\sqrt{2\pi}}\exp(-\frac{u^2}{2})
+(1-u^2)\text{erfc}\Big(\frac{u}{\sqrt{2}}\Big) \nonumber
\label{Recovering_Gardner}
\end{eqnarray}
and hence 
\begin{equation}
A_{\text{syn}}(u,\infty;c)=c+\frac{2u}{\sqrt{2\pi}}
    \exp\Big(-\frac{u^2}{2}\Big) \,.
\end{equation}
These are precisely the Gardner results with dilution \cite{Bo90} after
rescaling $u/\sqrt{2}\to u$. We remark that in this case, and also for
the $Q$-Ising type models  \cite{Mo97}, it is possible to 
rescale the optimal capacity as follows
\begin{equation}
\frac{\alpha^{{RS}}_{\text{syn}}(\kappa,c)}
{\alpha^{{RS}}_{\text{syn}}(\kappa,c=1)}=c+\frac{2u}{\sqrt{2\pi}
}\exp({-\frac{u^2}{2}})
\label{eq:rescale}
\end{equation}
with $c=\text{ erfc}(u/\sqrt{2})$. For the general BEG-perceptron
treated here such a scaling is not possible because the factor $\gamma$ 
appears both in the numerator and denominator of 
Eq.\eqref{capacity_RS_synchronous}. It is possible, however, to derive
the bound 
\begin{equation}
c-c\log c\leq \frac{\alpha_{RS}(\kappa,a,c)}{\alpha_{RS}(\kappa,a,c=1)}\leq 
c+\frac{2u}{\sqrt{2\pi}}\exp({-\frac{u^2}{2}})
\end{equation}
for $ 0\leq a\leq 1$ with $c$ and $u$ 
related through $c=~\text{ erfc}(u/\sqrt{2})$. These bounds are shown in
Fig.~1 as the broken line (lower bound) and the dashed-dotted line
(upper bound). Although the dependence on the dilution and other 
parameters is not that simple, we do find that
the dependence on the embedding stability parameter $\kappa$ is rather
weak.
\subsection{Asynchronous dilution}
In this case   $c_J$ is different from $c_K$ allowing us to study the
relative influence of the two sets of couplings. An analogous
calculation as the one in subsection A can be done leading to the
following result for the optimal capacity 
\begin{figure}[ht]
\begin{center}
\includegraphics[width=0.30\textwidth,height=0.25\textheight]
{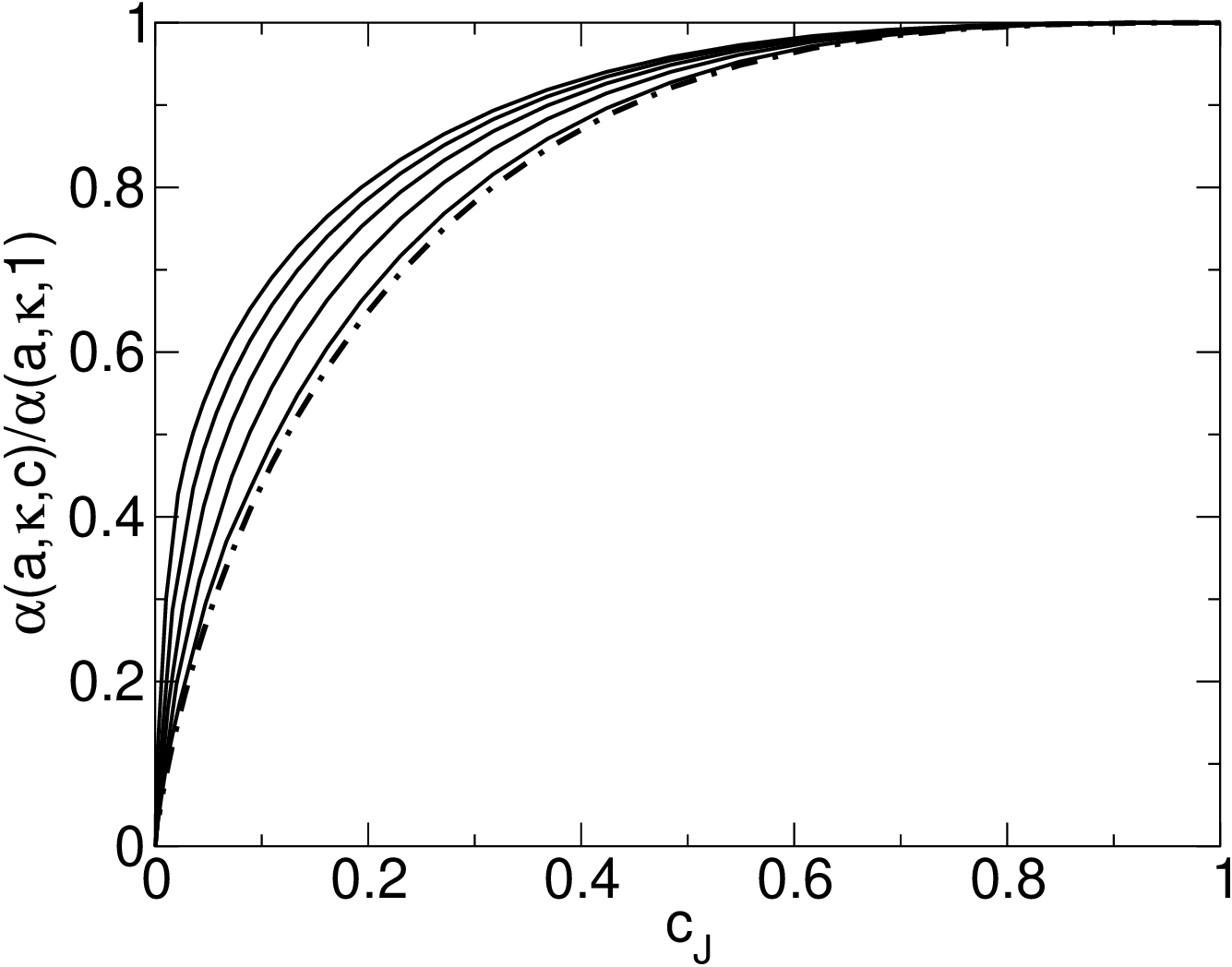}
\\ 
\vspace{1cm}
\includegraphics[width=0.30\textwidth,height=0.25\textheight]
{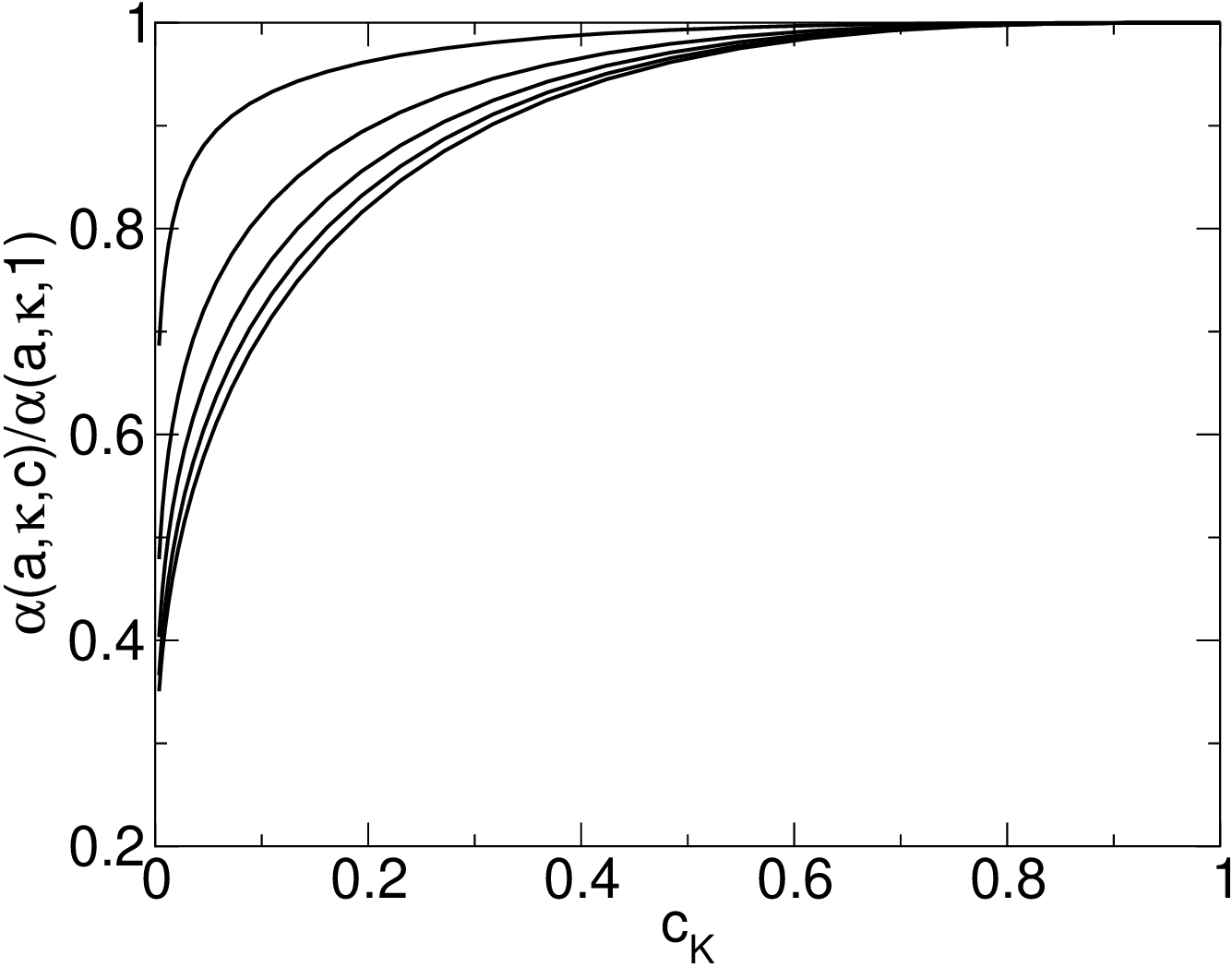}
\caption{Optimal capacity for asynchronous dilution in the
BEG-perceptron. Top figure:  the normalized optimal capacity as a
function of $c_J$ for $c_K=1$, $\kappa=0.0$ and activity 
$a=0.1,0.3,0.5,0.7, 0.9$ from top to bottom. The dashed-dotted line is for
$a=1$. Bottom figure: the
normalized capacity as a function of  $c_K$ for $c_J=1$,
$\kappa=0.0$ and pattern activity $a=0.1,0.3,0.5,0.7,0.9$
from bottom to top.}
\end{center}
\end{figure}

\begin{eqnarray}
&&\alpha_{\text{asyn}}^{{RS}}(a,\kappa,c_J,c_K)=
            \nonumber\\
&&\underset{u_J,u_K,L,\gamma}{\text{extr}}
  \frac{A_{\text{asyn}}(u_J;c_J)
    +\frac{1}{\gamma}{A_{\text{asyn}}(u_K;c_K)}}{g(\gamma,L;a,\kappa)}
\label{capacity_RS_asynchronous}
\end{eqnarray}
with
\begin{eqnarray}
&&A_{\text{asyn}}(u;c)= cu^2+\sqrt{\frac{2}{\pi}}u 
\exp({-\frac{u^2}{2}})+2(1-u^2)H(u) \nonumber\\ \\
&&H(u)= \frac{1}{\sqrt{2 \pi}} \int_u^\infty dx \exp(-\frac{x^2}{2})\,.
\end{eqnarray}
To simplify notation we denote both $u_J$ and $u_K$ as $u$ in the sequel
since there should be no confusion possible.

Stationarity with respect to $u$ leads to $c=\text{ erfc}(u/\sqrt{2})$.
We remark that when we take the dilution averages to be equal, i.e.,
$c_J=c_K\equiv c$ the dependence on the dilution in 
\eqref{capacity_RS_asynchronous}
factorizes and we simply get an expression equivalent to \eqref{eq:rescale}
\begin{equation}
\frac{\alpha^{{RS}}_{\text{asyn}}(a,\kappa,c)}
     {\alpha^{{RS}}_{\text{asyn}}(a,\kappa,c=1)}=
        A_{\text{asyn}}(u;c) 
\end{equation}
for any value of the pattern activity $a$ and stability constant $\kappa$.

In order to understand the role of the different couplings 
in the learning process, we cut them independently and study the 
influence with varying activity. The results are presented in Fig.~2.
We plot the  optimal capacity normalized by its value for  
no dilution, $\alpha(a,\kappa,c)/\alpha(a,\kappa,1)$ as a function of
the dilution of the $J$-couplings, $c_J$, for $c_K=1$, $\kappa=0$ and 
several values of the activity $a$ (top) and, analogously (bottom) as a
function of the dilution of the $K$-couplings.
We find that when diluting the $J$-couplings, referring to the ternary
patterns, and keeping all the $K$-couplings, related to the 
 active patterns, the normalized  capacity decreases 
as a function of the activity obtaining the Gardner result for $a=1$. 
When doing the reverse, the normalized capacity increases as a function
of the activity. Moreover, the network is more robust against
$K$-dilution, especially for large activities. This seems to be quite 
natural since large activities means many active states $\pm 1$ such
that cutting active patterns becomes relatively less important.

\section{Distribution of couplings}
\label{sec:distribution}
We study the distribution of couplings $\rho(J,K)$ inside $V$ in analogy
with \cite{Bo90}. This probability distribution can be splitted into two
parts, the first one involving  the $(1-c)N$ inactive couplings and 
the second one, $\rho_r(J,K)$, representing the remaining $cN$ active 
couplings. Obviously, the first set of couplings is delta distributed so that
we can write
\begin{equation}
    \rho(J,K)=(1-c)\delta(J)\delta(K)+\rho_r(J,K)
\end{equation}
where the second set of couplings satisfies
\begin{widetext}
\begin{eqnarray}
\rho_r(J,K)&=&\lim_{N\to\infty}\Odiss\frac{1}{V}\prod_{j=1}^N
\underset{ c_j, J_j, K_j}{\text{tr}}
  \delta\Big(\sum_{i=1}^N c_{i}J_i^2-cN\Big)
  \delta\Big(\sum_{i=1}^N c_{i}K_i^2-cN\Big)\times\nonumber\\
&&\delta\Big(\sum_{i=1}^N c_{i}-cN\Big)\prod_{\mu=1}^p
\chi_{\xi^\mu_0}(h^\mu,\theta^\mu;\kappa)\delta(J_1-J)
\delta(K_1-K)\delta_{c_1,1}\Cdiss \,.
\end{eqnarray}
\end{widetext}
In order to compute $\rho_r(J,K)$ we follow \cite{Bo90} by introducing 
replicas allowing us to lift the volume $V$ to the numerator. 
The calculations are standard but tedious. Evaluating the 
expression within the RS approximation we get for synchronous dilution
\begin{widetext}
\begin{equation}
\rho_{r, \text{syn}}(J,K)=\frac{\gamma A_{\text{syn}}(u,\gamma;c)}
             {2\pi c(1+\gamma)}
\exp \left[{-\frac{\gamma A_{\text{syn}}(u,\gamma;c)}{2c(1+\gamma)}
    \big( J^2+ K^2\big) }\right]
\Theta\left(\frac{\gamma A_{\text{syn}}(u,\gamma;c)}{c(1+\gamma)}
\Big(J^2+\frac{K^2}{\gamma}\Big)-u^2\right) \,.
\label{distribution_couplings_syn}
\end{equation}
\end{widetext}
This distribution is a two-dimensional Gaussian from which the middle
section has been cut out, as represented by the Heaviside function. This
gap has an ellipsoidal shape because of the scaling factor $1/\gamma$ 
accompanying the $K^2$ in the argument. It increases with increasing
dilution to reach its maximum when $c$ tends to zero. In the limit 
$\gamma\to\infty$  this distribution reduces to
\begin{widetext}
\begin{eqnarray}
\lim_{\gamma\to\infty}\rho_{r,\text{syn}}(J,K)=
    \frac{ A_{\text{syn}}(u,\infty;c)}{2\pi c}
\exp\left[{-\frac{ A_{\text{syn}}(u,\infty;c)}{2c}\big( J^2+ K^2\big)
     }\right]
\Theta\left(\frac{ A_{\text{syn}}(u,\infty;c)}{c}J^2-u^2\right)\,.
\end{eqnarray}
\end{widetext}
We remark that this distribution is different from the one obtained in
the Gardner case (i.e., the $a\to1$ limit) because, although the $K$
couplings do not play any role for $\gamma\to\infty$ the spherical 
constraint is still present, no matter what the value of $a$ is.

It is interesting to determine how  this probability distribution
behaves in the case of no dilution. Then, the distribution 
\eqref{distribution_couplings_syn} 
for the couplings becomes Gaussian without a gap, viz.
\begin{equation}
 \rho_{r,\text{syn}}(J,K;c=1)=
    \frac{1}{2\pi}\exp\left({-\frac{J^2}{2}-\frac{K^2}{2}}\right)\, .
\label{distribution_fc}
\end{equation}
This result is intuitively meaningful since the couplings 
are forced to obey only the spherical constraint without any
restriction coming from the dilution variable. Therefore, we find
back the probability distribution for the couplings of the fully connected
BEG-perceptron.

For asynchronous dilution a similar treatment can be pursued and we find
that the probability distribution for the couplings factorizes 
\begin{widetext}
\begin{eqnarray}
\rho_{r, \text{asyn}}(J)
   &=&\frac{1}{\sqrt{2\pi}}\sqrt{\frac{A_{\text{asyn}}(c_J,u_J)}{c_J}}
      \exp\left({-\frac{A_{\text{asyn}}(c_J,u_J)}{2c_J} J^2}\right)
    \Theta\left(|J|-u_J\sqrt{\frac{c_J}{A_{\text{asyn}}(c_J,u_J)}}\,\right)
	   \\
\rho_{r,\text{asyn}}(K)
   &=&\frac{1}{\sqrt{2\pi}}\sqrt{\frac{A_{\text{asyn}}(c_K,u_K)}{c_K}}
       \exp\left({-\frac{A_{\text{asyn}}(c_K,u_K)}{2c_K} K^2}\right)
      \Theta\left(|K|- u_K\sqrt{\frac{c_K}{A_{\text{asyn}}(c_K,u_K)}}\right)\, .
\end{eqnarray}
\end{widetext}
These distributions are of a similar nature as the one for the standard 
diluted perceptron case \cite{Bo90}.

\section{de Almeida-Thouless stability}
\label{sec:stability}
Finally, we are  interested in studying the local stability of
the obtained solutions against RS fluctuations following 
\cite{AT78,MPV,Pe03}. From the work on the non-diluted BEG-perceptron 
\cite{Pe03} we recall that in that case the solutions are unstable only 
for small  activities and very small embedding constants $\kappa$. 
Furthermore, we know that, in general, there are four transverse 
eigenvalues. In the case of asynchronous dilution these eigenvalues are 
given by the roots of the fourth degree characteristic polynomial
\begin{equation}
P(\lambda)=
\begin{vmatrix}
        \Delta_q-\lambda&\Delta_c&c_J&0\\
         \Delta_c&\Delta_r-\lambda&0&c_K\\
          c_J&0&\Delta_{\widehat{q}}-\lambda&0\\
          0&c_K&0&\Delta_{\widehat{r}}-\lambda
\end{vmatrix}=0
\end{equation}
where $\Delta_{\widehat{q}}$ and $\Delta_{\widehat{r}}$ read
\begin{eqnarray}
\Delta_{\widehat{q}}&=&
\int D(x)\Big[\frac{1}{\widehat{q}}\frac{\partial^2}{\partial x^2}
\log[1]_\dag(x,E,\widehat{q},\psi_J)\Big]^2\label{eq:stability1}\\
\Delta_{\widehat{r}}&=&
\int D(x)\Big[\frac{1}{\widehat{r}}\frac{\partial^2}{\partial x^2}
\log[1]_\dag(x,F,\widehat{r},\psi_K)\Big]^2\label{eq:stability2}
\label{b}
\end{eqnarray}
with $E$, $F$, $\phi_J$ and $\psi_K$ the conjugate variables appearing
in the integral representations of the constraints, $\hat q$ and $\hat r$ 
the conjugate variables of the order parameters $q$ and $r$, and with
the short-hand notation
\begin{eqnarray}
[1]_\dag(x,a,b,d)&=&1+\sqrt{\frac{2\pi}{a-b}}
\exp\Big[-\frac{bx^2}{2(a-b)}-\frac{d}{2}\Big]\, . \nonumber
\end{eqnarray}
Similar expressions can be written down for $\Delta_{q}$, $\Delta_{r}$ 
and $\Delta_{c}$ but they are not needed for the argumentation. Indeed,
it is straightforward to check that as soon as dilution is allowed  
the solution becomes unstable in the saturation limit  $q\to1$. The first 
derivative of $[1]_\dag(x,a,b,d)$ has a jump at $x=u$ proportional 
to $u$, leading to a  dirac delta contribution in the second derivative. 
The square in \eqref{eq:stability1} and \eqref{eq:stability2} forces the
replicon eigenvalue to go to $+\infty$,  similarly to what happens for
the standard perceptron model as explained in \cite{Bo94}.  
When $u=0$, i.e. in the absence of dilution, there is no such delta 
contribution and we find back the results of \cite{Pe03}.
The same reasoning holds for synchronous dilution.

\section{Conclusions}
\label{sec:conclusions}
In this work we have studied annealed dilution in the 
BEG perceptron model. Two types of dilution have been
discussed, the first one being synchronous dilution, i.e., simultaneous 
dilution of some of the couplings referring to the ternary patterns 
themselves and  some of the couplings related to  
the active patterns, the second one being dilution of both these types of
couplings independently, so-called asynchronous dilution. 
We have obtained an analytic formula for the replica  symmetric Gardner
optimal capacity. 
For synchronous dilution we see that different regions of activities lead 
to different results.
For small activities $a \leq 0.5$  the optimal capacity strongly increases
for decreasing dilution but normalizing these  results by
its value for no dilution, the lines for different activities collapse.
For large  activities $a \geq 0.6$  the optimal storage capacity
is only weakly
dependent on the activity but the results for the normalized
optimal capacity do not  collapse. Furthermore, we see that the network is
robust against synchronous dilution for non-sparse coding, i.e.,
for activities ranging in the interval $[\,0.2, 1.0\,]$.
For asynchronous dilution we find that diluting only the $J$-couplings,  
the normalized optimal capacity decreases 
as a function of the activity obtaining the Gardner result for $a=1$. 
When diluting the $K$-couplings, the normalized optimal capacity increases
as a function of the activity. Moreover, the network is more robust against
$K$-dilution, especially for large activities.   
Since the effects of dilution are of the same order as those in the $Q=3$-Ising  
model, these results also confirm the better retrieval 
properties found before for the BEG model. 

We have studied the stability of the RS solution against
RS breaking fluctuations by      
generalizing the de Almeida-Thouless analysis. We find that as soon as
there is dilution the results are unstable.

\begin{acknowledgments}
We are indebted to Jort van Mourik and Nikos Skantzos for critical 
and informative discussions. This work has been supported  by the Fund of 
Scientific Research, Flanders-Belgium.
\end{acknowledgments}

\section*{APPENDIX A}
In this appendix we outline the main steps in the calculation of the
entropy per neuron \eqref{entropybis}-\eqref{power_volume}.

After defining the  order parameters
\begin{eqnarray}  
   L^\alpha&=&\frac{1}{\sqrt{cN}}\sum_{j=1}^N c^\alpha_jK^\alpha_j,
   \quad \forall \alpha\\
   q_{\alpha\beta}&=&\frac{1}{cN}
             \sum_{j=1}^N c^\alpha_j J^\alpha_jc^\beta_jJ^\beta_j,
	\quad  \alpha < \beta     \\
   r_{\alpha\beta}&=&\frac{1}{cN}
               \sum_{j=1}^N c^\alpha_j K^\alpha_jc^\beta_jK^\beta_j, 
       \quad  \alpha < \beta
\label{order_parameters}
\end{eqnarray}
introducing the conjugate order parameters 
$\widehat L^\alpha, \widehat q_{\alpha\beta},    
\widehat r_{\alpha\beta}$, and enforcing the constraints 
\eqref{constraint1} and \eqref{constraint23} using the Lagrange 
multipliers $E^\alpha$, $F^\alpha$ and $\widehat{\psi}^\alpha$, 
we write $\odis V^n\cdis$ as the following integral 
\begin{widetext}
\begin{eqnarray}
\odis V^n\cdis &=&\int \Big[\prod_{\alpha<\beta}
\frac{d q_{\alpha\beta}d\widehat{q}_{\alpha\beta}}{2\pi i/cN}
\frac{d r_{\alpha\beta}d\widehat{r}_{\alpha\beta}}{2\pi i/cN}\Big]
\Big[\prod_{\alpha=1}^n\frac{d L^{\alpha}d\widehat{L}^{\alpha}}
                    {2\pi /\sqrt{cN}}\frac{dE^\alpha}{4\pi i}
     \frac{dF^\alpha}{4\pi i}\frac{d\widehat{\psi}^\alpha}{4\pi i}\Big]
    \exp[{N(G_1+G_2+G_3)}]
\end{eqnarray}
where we have defined the  functions
\begin{eqnarray}
G_1&=&\alpha\log\int\Big[\prod_{\alpha=1}^n
\frac{dh^\alpha d\widehat{h}^\alpha}{2\pi}\Big]
\Big[\prod_{\alpha=1}^n\frac{d\theta^\alpha 
d\widehat{\theta}^\alpha}{2\pi}\Big]
\exp\Big[[i\sum_{\alpha=1}^n\Big( h^\alpha \widehat{h}^\alpha
+\theta^\alpha \widehat{\theta}^\alpha\Big)
-ia \sum_{\alpha=1}^n\widehat{\theta}^\alpha L^\alpha\nonumber\\
&&-\frac{a}{2} \sum_{\alpha,\beta=1}^n
\widehat{h}^\alpha\widehat{h}^\beta q_{\alpha\beta}
-\frac{a(1-a)}{2} 
\sum_{\alpha,\beta=1}^n\widehat{\theta}^\alpha\widehat{\theta}^\beta
 r_{\alpha\beta}\Big\}\Big]\Odis\prod_{\alpha=1}^n
\chi_{\xi_0}(h_\alpha,\theta_\alpha;\kappa)\Cdis_{\xi_0}\\
G_2&=&\log\prod_{\alpha=1}^n\text{tr}_{\{c^\alpha,J^\alpha,K^\alpha\}}
\exp\Big[-\sum_{\alpha<\beta}\widehat{q}_{\alpha\beta} 
c^\alpha J^{\alpha}c^\beta J^{\beta}-
  \sum_{\alpha<\beta}\widehat{r}_{\alpha\beta}
 c^\alpha K^{\alpha}c^\beta K^{\beta}\nonumber\\
&&-\frac{1}{2}\sum_{\alpha=1}^n E^\alpha c^\alpha(J^\alpha)^2
-\frac{1}{2}\sum_{\alpha=1}^n F^\alpha c^\alpha(K^\alpha)^2
-\frac{1}{2}\sum_{\alpha=1}^n \widehat{\psi}^\alpha c^\alpha\Big]\\
G_{3}&=&c\sum_{\alpha<\beta}\widehat{q}_{\alpha\beta}q_{\alpha\beta}
+c\sum_{\alpha<\beta}\widehat{r}_{\alpha\beta}r_{\alpha\beta}
+\frac{c}{2}\sum_{\alpha=1}^n\Big[E^\alpha+F^\alpha 
+\widehat{\psi}^\alpha\Big]\,.
\end{eqnarray}
\end{widetext}
We have already used that $\widehat{L}^\alpha=0$, $\forall\alpha$ at the 
saddle-point. 
In the thermodynamic limit $N \to \infty$ the entropy is 
evaluated at the saddle-point for the order parameters 
\eqref{order_parameters}, the  conjugate ones and the Lagrange multipliers
 $E^\alpha$, $F^\alpha$ and $\widehat{\psi}^\alpha$.
Using the RS ansatz for the order parameters 
\begin{eqnarray}
&& L^\alpha=L\quad E^\alpha=E\quad 
F^\alpha=F\quad \widehat{\psi}^\alpha=\widehat{\psi} \\
&&q_{\alpha\beta}=q \quad r_{\alpha\beta}=r\quad
\quad\widehat{q}_{\alpha\beta}=\widehat{q} 
\quad \widehat{r}_{\alpha\beta}=\widehat{r}
\end{eqnarray}
the functions $G_1,G_2,G_3$ can be simplified further and the entropy can be    
written as
\begin{widetext}
\begin{equation}  
v=-\frac{c}{2}\widehat{q}q-\frac{c}{2}\widehat{r}r+
\frac{c}{2}\Big[E+F +\widehat{\psi}\Big]
+\int_{-\infty}^\infty{\cal D}(x,y)\log[1]_\dag(x,y)
+\alpha\int_{-\infty}^\infty{\cal D}(h_0,\theta_0-l)
\odis\log [1]^{\xi}_{\star}(h_0,\theta_0) \cdis
\label{entropy_RS}
\end{equation}
with  the short-hand notations
\begin{eqnarray}
~[1]_\dag(x,y)&=&1+\frac{2\pi}{\sqrt{(E-\widehat{q})(F-\widehat{r})}}
\exp\left[{-\frac{x^2\widehat{q}}{2(E-\widehat{q})}
-\frac{y^2\widehat{r}}{2(F-\widehat{r})}-\frac{\psi}{2}}\right]
      \label{function1}\\
~[1]^{\xi}_{\star}(h_0,\theta_0)&=&\int_{\Omega_{\xi}}
   \frac{dh}{\sqrt{2\pi(1-q)}} \frac{d\theta}{\sqrt{2\pi (1-r)}}
  \exp\left[{-\frac{(h-h_0)^2}{2(1-q)}
    -\frac{(\theta-\theta_0)^2}{2(1-r)}}\right]\label{function2}
\end{eqnarray}
\end{widetext}
where the integral in \eqref{function2}
is restricted to the region $\Omega_{\xi}$ given by the 
characteristic function $\chi_{\xi}(h\sqrt{a},\theta\sqrt{a(1-a)};\kappa)$
 defined in \eqref{charateristic_function}. The RS Gardner optimal 
capacity  is then  reached when $q,r$  go to 1.

At this point we have two choices to proceed. Either we 
solve numerically the saddle-point equations or we do 
an asymptotic expansion in the limit $q,r\to1$ in the 
entropy \eqref{entropy_RS} (or equivalently in the saddle-point
equations for the parameters). The first approach has the 
advantage that we can study $\alpha$ as a function
of $(q,r)$. But, since we are only interested in the 
optimal capacity, we opt for the asymptotic expansion.
Since the limits $q\to 1$ and $r \to 1$ must be taken 
simultaneously, we introduce a factor $\gamma$ such that 
$(1-r)=\gamma(1-q)$. Then, a simple inspection  of
 the function \eqref{function2}  appearing in the expression 
of the entropy \eqref{entropy_RS} suggests that in the 
limit $q\to1$ this function will diverge as $(1-q)^{-1}$.
Since this function is coupled to the capacity and we expect 
non-trivial results, the other terms in the entropy also have
to diverge in such a way. This implies, for instance, that for 
the function $[1]_\dag(x,y)$ the terms 
$\widehat{q}/(E-\widehat{q})$, $\widehat{\psi}$ and  
$\widehat{r}/(F-\widehat{r})$ appearing in its argument 
have to go to infinity as $(1-q)^{-1}$. The precise 
coefficients in front of this divergence are given 
by the saddle-point equations of the conjugated
order-parameters.
Performimg this asymptotic expansion explicitly leads to the result
\eqref{capacity_RS_synchronous} in Section \ref{sec:Gardner}.


\begin{thebibliography}{99}
\bibitem{BEG} M. Blume, V.J. Emery, and R.B. Griffiths, Phys. Rev. A 
{\bf 4}, 1071 (1971); M. Blume, Phys. Rev. {\bf 141}, 517 (1966); 
H.W. Capel, Physica (Amsterdam) {\bf 32}, 966 (1966).
\bibitem{Do00} D.R. Dominguez Carreta and E. Korutcheva, Phys. Rev. E 
 {\bf 62}, 2620 (2000).
\bibitem{Ve03} D. Boll\'e and T. Verbeiren, J. Phys. A: Math. Gen 
{\bf 36}, 295, 2003.
\bibitem{BRS94} D. Boll\'e, H. Rieger, and G.M. Shim, J. Phys. A: Math. Gen
{\bf 27}, 3411, 1994.
\bibitem{Ga87} E. Gardner, J. Phys. A: Math. Gen. {\bf 21}, 257 (1988).
\bibitem{Pe03} D. Boll\'e, I. P\'erez Castillo, and G. M. Shim,
Phys. Rev. E {\bf 67}, 036113, 2003.
\bibitem{MKB91}S. Mertens, H. M. K\"ohler, and S. B\"os,
   J. Phys. A {\bf 24}, 4941 (1991).
\bibitem{BDM91}D. Boll\'e, P. Dupont, and J. van Mourik,
   Europhys. Lett. {\bf 15}, 893 (1991).
\bibitem{GK94}F. Gerl and U. Krey, J. Phys. A {\bf 27}, 7353 (1994).
\bibitem{GBK} F. Gerl, K. Bauer, and U. Krey,  Z. Phys. B {\bf 88}, 
   339 (1992).
\bibitem{De87} B. Derrida, E. Gardner, and A. Zippelius, Europhys. Lett
  {\bf 4}, 167 (1987).
\bibitem{Ga89} E. Gardner, J. Phys. A: Math. Gen. {\bf 22}, 1969 (1989).
\bibitem{Co03} B. Wemmenhove and A. C. C. Coolen  J. Phys. A: Math. Gen. 
{\bf 63}, 9617 (2003).
\bibitem{Pe03b} I. P\'erez Castillo and N. Skantzos, cond-mat/0309655
\bibitem{Bo90} M. Bouten, A. Engel, A. Komoda, and R. Serneels,
 J. Phys. A: Math. Gen {\bf 23},4643 (1990).
\bibitem{Mo97} D. Boll\'e and J. van Mourik, J. Phys. A: Math. Gen.,
   {\bf 27}, 1151 (1997).
\bibitem{MPV} M. M\'ezard, G. Parisi, and M.A. Virasoro, {\em Spin Glass 
Theory and Beyond}, Singapore, World Scientific (1987).
\bibitem{T88} M.V. Tsodyks, Europhys. Lett. {\bf 7}, 203 (1988). 
\bibitem{PV89} C.J. Perez-Vicente, Europhys. Lett. {\bf 10}, 621 (1989). 
\bibitem{H89} H. Horner, Z. Phys. B {\bf 75}, 133 (1989).
\bibitem{AT78} J. R. de Almeida and D. Thouless, J. Phys. A: Math. Gen.,
 {\bf 11}, 983 (1978).
\bibitem{Bo94} M. Bouten, J. Phys. A: Math. Gen., {\bf 27}, 6021 (1994).

\end{thebibliography}
\end{document}